\documentclass[12pt]{article}
\usepackage{graphicx}
\usepackage{latexsym,amsfonts, amssymb,epsfig}
\usepackage[russian,english]{babel}

\begin{document}

\begin{center}
{\Large\bf Compacton-like solutions of the hydrodynamic system
describing relaxing media}

\vspace{10mm}

{\it {\large\bf V.A. Vladimirov}
\\
 \vspace{5mm}

  University of Science and Technology\\
Faculty of Applied Mathematics \\ Al Mickiewicza 30, 30-059 Krakow, Poland \\ E-mail vsan@rambler.ru \\[2ex] }

\vspace{3mm}

(Submitted to ROMP)

 \end{center}

\vspace{10mm}

{ \footnotesize {\bf Abstract}We show the existence of  a
compacton-like solutions within the relaxing
hyd\-ro\-dy\-na\-mic-type model and perform numerical study of
attracting features of these solutions

}

 \vspace{3mm}


\vspace{3mm}


\section{ Introduction }

 In this paper there are studied solutions to evolutionary
 equations, describing wave patterns with compact support.
 Different kinds of wave patterns play  key rules in natural
 processes. They occur in  nonlinear transport
 phenomena \cite{Shuster}, serve as a channels of  information transfer in
animate systems \cite{Davydov}, and very often assure stability of
some dynamical
 processes \cite{spindeton1,spindeton2}.

 One of the most advanced mathematical theory dealing with wave patterns'  formation and evolution is
 the soliton theory \cite{Dodd}. The origin of this theory goes back
 to Scott Russell's description of the solitary wave movement in the
 surface of  channel filled with water. It was the ability of the wave to move quite a long
 distance without any change of shape which stroke the imagination
 of the first chronicler of this phenomenon. In 1895 Korteveg
 and  de Vries put forward their famous equation
 \begin{equation}\label{KdV}
u_t+\beta\,u\,u_x+u_{xxx}=0,
 \end{equation}
describing long waves' evolution on the surface of a shallow water.
They also obtained the analytical solution to this equation,
corresponding to the solitary wave:
\begin{equation}\label{KdVsol}
u=\frac{12\,a^2}{\beta}\,sech^2{\left[a(x-4\,a^2 \,t)\right]}.
\end{equation}
Both the already mentioned report by Scott Russell as well as the
model suggested to explain his observation did not involve a proper
impact till the middle of 60-th of the XX century when there have
been established a number of outstanding features of equation
(\ref{KdV}) finally becoming aware as the consequences of its
complete integrability \cite{Dodd}.

In recent years ago there have been discovered another type of
solitary waves referred to as {\it compactons} \cite{Ros_93}. These
solutions inherit main features of solitons, but differ from them in
one point: their supports are compact.

A big progress is actually observed in studying compactons and their
properties, yet most papers dealing with this subject are concerned
with the compactons being the solutions to either completely
integrable equations, or those which produce a completely integrable
ones when being reduced onto subset of a traveling wave (TW)
solutions \cite{OlvRosenau,LiOlv,Ros_96}.

In this paper  compacton-like solutions to  the hydrodynamic-type
model taking account of the effects of temporal non-locality are
considered. Being of dissipative type, this model is obviously
non-Hamiltonian. As a consequence, compactons are shown to exist
merely for selected values of the parameters. In spite of such
restriction, existence of this type of solution in significant for
several reasons. Firstly, the mere existence of this solutions is
connected with the presence of relaxing effects and rather cannot be
manifested in any local hydrodynamic model. Secondly, solutions of
this type manifest some attractive features and can be treated as
some universal mechanism of the energy transfer in media with
internal structure leading to the given type of the
hydrodynamic-type modeling system.

The structure of the paper is following. In section~2 we give a
geometric insight into the soliton and compacton TW solutions,
revealing the mechanism of appearance of generalized solutions with
compact supports. In  section~3 we introduce the modeling system and
show that compacton-like solutions do exist among the set of TW
solutions. In  section~4 we perform the numerical investigations of
the modeling system based on the Godunov method and show that
compacton-like solutions manifest attractive features.

\section{Solitons and compactons from the geometric viewpoint}

\begin{figure}
\centering \includegraphics[width=2.5 in, height=2.0 in
]{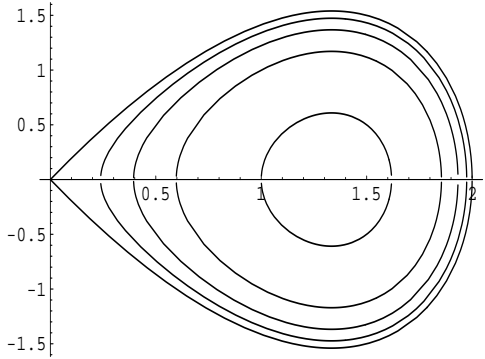} \caption{Level curves of the Hamiltonian function of
system (\ref{ODE_solit}) representing periodic solutions and
limiting to them homoclinic solution} \label{hclKdV}
\end{figure}

Let us discuss how the solitary wave solution to (\ref{KdV})  can be
obtained. Since the function $u(\cdot)$ in the formula
(\ref{KdVsol}) depend on the specific combination of the independent
variables, we can use for this purpose the ansatz $u(t,\,x)=U(\xi),
\,\,\,\mbox{with}\,\,\, \xi=x-V\,t.$ Inserting this ansatz into
equation (\ref{KdV}) we  get, after one integration, following
system of ODEs:
\begin{eqnarray}
\dot U(\xi)=-W(\xi),\label{ODE_solit} \\
\dot W(\xi)=\frac{\beta}{2}\,U(\xi)\left(U(\xi)-
\frac{2\,V}{\beta}\right). \nonumber
 \end{eqnarray}
System (\ref{ODE_solit}) is a Hamiltonian system describing by  the
Hamilton function
\[
H=\frac{1}{2}\,\left(W^2+\frac{\beta}{3}\,U^3-V\,U^2  \right).
\]
Every solutions of (\ref{ODE_solit}) can be identified with some
level curve $H=C$. Solution (\ref{KdVsol}) corresponds  to the value
$C=0$ and is represented by the homoclinic trajectory shown in
Fig.~\ref{hclKdV} (the only  trajectory  in the right half-plane
going through the origin). Since the origin is an equilibrium point
of system (\ref{ODE_solit}) and penetration of the hohoclinic loop
takes an infinite "time" then the beginning of this trajectory
corresponds to $\xi=-\infty$ while its end - to $\xi=+\infty$. This
assertion is equivalent to the statement that solution
(\ref{KdVsol}) is nonzero for all finite values of the arguments.

Now let us discuss the geometric structure of compactons. For this
purpose we restore to the original equation which is a nonlinear
generalization to classical Korteveg - de Vries equation
\cite{Ros_93}:
\begin{equation}\label{Compeq}
u_t+\alpha\,u^m\,u_x+\beta\left(u^n\right)_{xxx}=0.
\end{equation}
Like in the case of equation (\ref{KdV}), we look for the TW
solutions $u(t,\,x)=U(\xi)$, where $\xi=x-V\,t$. Inserting this
ansatz into (\ref{Compeq}) we obtain, after one integration, the
following dynamical system:
\begin{eqnarray}
\frac{d\, U}{d\,T}=-n\,\beta\,U^{2}\,W,\label{comp_ODE} \\
\frac{d\,
W}{d\,T}=U^{3-n}\left[-V\,U+\frac{\alpha}{m+1}U^{m+1}-n\,\beta\,U^{n-2}W\right],
\nonumber
 \end{eqnarray}
 where $\frac{d}{d\,T}=n\,\beta\,U^{2}\frac{d}{d\,\xi}$.
All the trajectories of this system are given by its first integral
\[
\frac{\alpha}{(m+1)\,(5+m-n)}\,U^{5+m-n}-\frac{V}{5-n}\,U^{5-n}+\frac{\beta\,n}{2}\left(U\,W
\right)^2=H=const.
\]

\begin{figure}
\centering \includegraphics[width=2.5 in, height=2.0 in
]{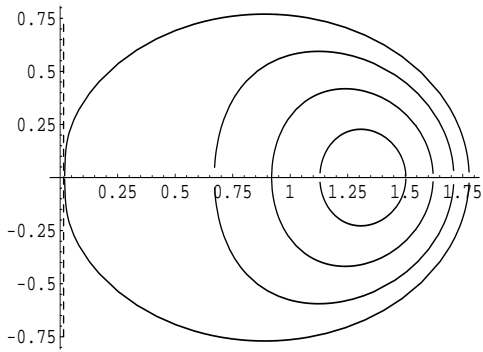} \caption{Level curves of the Hamiltonian function of
system (\ref{comp_ODE}). Dashed line indicates the set of singular
points $U=0$.} \label{Ros}
\end{figure}

Phase portrait of system (\ref{comp_ODE})  shown in Fig.~\ref{Ros}
are similar to some extent to that corresponding to system
(\ref{ODE_solit}). Yet the critical point $U=W=0$ of system
 (\ref{comp_ODE}) lies on the line of singular points $U=0$. And this
 implies that modulus of the tangent vector field along the homoclinic
 trajectory is bounded from below by a positive constant. Consequently
 the homoclinic trajectory is penetrated in
 a finite time and the corresponding generalized solution to the initial system
 ( \ref{Compeq}) is the compound of a function
 corresponding to the homoclinic loop (which now has a compact
 support) and zero solution corresponding to the rest point $U=W=0$.
 In case when $m=1, \,\,\,\beta=1/2$ and $n=2$ such solution has the following
 analytical representation \cite{Ros_93}:
\begin{equation}\label{companalsol}
u=\left\{\begin{array}{c}  \frac{8\,V}{3\,\alpha}\,\cos^2{\sqrt{\frac{\alpha}{4}}\,\xi} \,\,\,
\mbox{when}\,\,\, |\xi|<\frac{\pi}{\sqrt{\alpha}}, \\
0  \,\,\,\mbox{when}\,\,\, |\xi|\geq
\frac{\pi}{\sqrt{\alpha}}.\end{array} \right.
\end{equation}

It is quite obvious that similar mechanism of creating the
compacton-like solutions can be realized in case of  non-Hamiltonian
system as well, but in contrast to the Hamiltonian systems, the
homoclinic solution is no more expected to form a one-parameter
family like this is the case with solution (\ref{companalsol}). In
fact, in the modeling system described in the following section,
homoclinic solution appears as a result of a bifurcation following
the birth of the limit cycle and its further interaction with the
nearby saddle point.

Let us note in conclusion that  we do not distinguish solutions
having the compact supports and those which can be made so by proper
change of variables. In particular, the solutions we deal with in
the following sections, are realized as compact perturbations
evolving in a self-similar mode on the background of the stationary
inhomogeneous solution of a system of PDEs.

\section{Relaxing hydrodynamic-type model and its qualitative investigations}

We consider the following system \cite{DanDanVlad}:
\begin{eqnarray}
u_t+p_x=\gamma, \nonumber \\
V_t-u_x=0, \label{Relhydro} \\
\tau\,p_t+\frac{\chi}{V^2}\,u_x=\frac{\kappa}{V}-p. \nonumber
\end{eqnarray}
Here $u$ is mass velocity, $V$ is specific volume, $p$ is pressure,
$\gamma$ is acceleration of the external force, $\kappa$ and
$\chi/\tau$ are squares of the equilibrium and "frozen" sound
velocities, respectively, $t$ is time, $x$ is mass (Lagrangean)
coordinate. The first two equations are convenient balance equations
for   momentum and mass, while the last one is the dynamical
equation of state, taking into account relaxing properties of the
media.

We perform the factorization \cite{Olv} of system (\ref{Relhydro})
(or, in other words, passage to an ODE system describing TW
solutions), using its symmetry properties summarized in the
following statement.

\vspace{3mm}

 {\bf Lemma 1. } System (\ref{Relhydro}) is invariant
with respect to one-parameter groups of transformations generated by
the infinitesimal operators
\begin{equation}\label{symgen}
\left\{\begin{array}{c} \hat X_0=\frac{\partial}{\partial\,t},
\qquad \hat X_1=\frac{\partial}{\partial\,x}, \\
\hat
X_2=p\,\frac{\partial}{\partial\,p}+x\,\frac{\partial}{\partial\,x}-V\,\frac{\partial}{\partial\,V}.
\end{array}\right.
\end{equation}

\vspace{3mm}

{\it Proof.} Invariance with respect to one parameter groups
generated by the operators $\hat X_0,\,\,\hat X_1$ is a direct
consequence of the fact that system (\ref{Relhydro}) does not depend
explicitly on $t\,\,\mbox{and}\,\,x$. Operator $\hat X_2$ is the
generator of scaling transformation
\[
u'=u,\,\,p'=e^\alpha\,p,\,\,V'=e^{-\alpha}\,V,\,\,t'=t\,\,\,
\mbox{and}\,\,\,x'=e^\alpha\,x.
\]
Invariance of system (\ref{Relhydro}) with respect to this
transformation is easily verified by direct substitution.

\vspace{2mm}

The above symmetry generators are composed on the following
combination:
\[
\hat Z=\frac{\partial}{\partial\,t}+\xi\left[
(x-x_0)\,\frac{\partial}{\partial\,x}+p\,\frac{\partial}{\partial\,p}-V\,\frac{\partial}{\partial\,V}
\right].
\]
It is obvious that operator $\hat Z$ belongs to the Lie algebra of
the symmetry group of (\ref{Relhydro}). Therefore expressing the old
variables in terms of four independent solutions of equation $\hat
Z\,J(t,\,x)=0,$ we gain the reduction of the initial system
\cite{Olv}. Solving the equivalent system
\[
\frac{d\,t}{1}=\frac{d\,(x_0-x)}{\xi(x_0-x)}=\frac{d\,p}{\xi\,p}=\frac{d\,V}{-\xi\,V}=\frac{d\,u}{0},
\]
we get the following ansatz, leading to reduction:
\begin{equation}\label{Ans_red}
u=U(\omega),\,\,p=\Pi(\omega)\,(x_0-x),\,\,V=R(\omega)/(x_0-x),\,\,\omega=\xi\,t+\log{\frac{x_0}{x_0-x}}.
\end{equation}
In fact, inserting  this ansatz into the second equation of system
(\ref{Relhydro}), we get the quadrature
\begin{equation}\label{quadr}
U\,=\xi R\,+const.
\end{equation}
and the following dynamical system:
\begin{eqnarray}\label{DS_rel}
\xi \Delta(R) R^{\prime}\,=-R\,\left[ \sigma R\Pi \,-\kappa \,+\tau
\xi
R\gamma\right]=F_1, \\
\xi \Delta(R) \Pi ^{\prime}\,=\xi \,\left\{ \xi R\,(R\Pi \,-\kappa
)\,+\chi \, (\Pi\,+\gamma )\right\}=F_2, \nonumber
\end{eqnarray}
where $(\cdot )^{\prime}\,=d\,(\cdot )\,/d\omega ,\quad \Delta(R)
\,=\tau (\xi R)^2-\chi,\,\quad \sigma \,=1\,+\tau \xi .$

In case when $\gamma<0,$ system (\ref{DS_rel}) has three stationary
points in the right half-plane. One of them, having the coordinates
$R_0\,=0,\quad \Pi _0\,=-\gamma,$ lies in the vertical coordinate
axis. Another one having the coordinates $ R_1\,=-\kappa \,/\gamma,
\quad \Pi _1\,=-\gamma  $ is the only stationary point lying in the
physical parameters range. The last one one having the coordinates
\[
R_{2}\,= \sqrt{\frac{\chi \,}{\tau \xi ^2 }},\quad \Pi
_{2}\,=\frac{\kappa -\tau \xi \gamma R_{2}}{\sigma R_{2}},
\]
lies on the line of singular points $\tau (\xi R)^2-\chi=0.$

As was announced earlier, we are looking for the homoclinic
trajectory arising as a result of a limit cycle destruction. So in
the first step we should assure the fulfillment of the Andronov-Hopf
theorem statements in some stationary point. Since the only good
candidate for this purpose is the point
$A\left(R_1,\,\,\Pi_1\right)$, we put the origin into this point by
making the following change of the coordinates $X=R-R_{1}, \,\,
Y=\Pi -\Pi_{1}$ which gives us the system
\begin{equation}\label{DS_rel_can}
  \xi \Delta(R)
  \left( \begin{array}{c}
  X \\
  Y
  \end{array} \right)'=
  \left[ \begin{array}{cc}
  -\kappa, & -R_1^2 \sigma \\
  \kappa \xi^2, & (\xi R_1)^2 +\chi \xi
  \end{array} \right]
  \left( \begin{array}{c}
  X \\
  Y
  \end{array} \right)\,+\,
  \left( \begin{array}{c}
  H_1 \\
  H_2
  \end{array} \right),
\end{equation}
where
$$
H_1\,=\, -\left (\Pi_1X^2\,+\,2\sigma R_1XY\,+\,\sigma X^2Y\right),
$$
$$
H_2\,=\,\xi^2\left(\Pi_1X^2\,+\,2R_1XY\,+\,X^2Y\right).
$$
A necessary condition for  the limit cycle appearance read as
follows \cite{Has}:
\begin{eqnarray}
  sp \hat M=0\,\,\Leftrightarrow\,\,(\xi R_1)^2 +\chi \xi =\kappa,
\label{sp_0}
\\
 det \hat M> 0\,\,\Leftrightarrow\,\,\Omega^2 =\kappa \xi \Delta(R_1) >0.
\label{det_pos}
\end{eqnarray}
where $\hat M$ is the linearization matrix of system
(\ref{DS_rel_can}).  The inequality (\ref{det_pos}) will be
fulfilled if $\xi<0$ and the coordinate $R_1$ lies inside the set
$(0,\,\,\sqrt{\chi /(\tau \xi^2)})$. Note that another option, i.e.
when $\xi>0$ and $\Delta>0$ is forbidden from physical reason
\cite{Landau_hdyn}. On view of that, the critical value of $\xi$ is
expressed by the formula \begin{equation}\label{xicr} \xi
_{cr}\,=-\frac{\chi \,+\sqrt{\chi ^2\,+4\kappa R_1^2}}{2R_1^2}.
\end{equation}

\vspace{2mm}
 {\bf Remark. }Note that as a by-product of inequalities
inequalities (\ref{sp_0}), (\ref{det_pos}) we get the following
relations:
\begin{equation}\label{ineqsig}
-1<\tau\,\xi\,<0.
\end{equation}

\vspace{2mm}

To accomplish the study of the Andronov-Hopf bifurcation, we are
going to calculate the real part of the first Floquet index $C_1$
\cite{Has}. For this purpose we use the transformation
\begin{equation}  y_1 =X, \qquad
y_2 =-\frac{\kappa}{  \Omega}\,X-\frac{\sigma\,R_1^2}{\Omega}\,Y,
\end{equation}
enabling to pass from system (\ref{DS_rel_can}) to the canonical one
having the following anti-diagonal   linearization  matrix
\[
  \hat M_{ij} =\Omega
  (\delta_{2i}\delta_{1j}-\delta_{1i}\delta_{2j}).
\]
For this representation formulae from \cite{Has,GH}, are directly
applied and, using them we obtain the  expression:
\[
  16\,R_1^2\,\Omega^2\,Re \ C_1 =-\kappa\,\left\{3 \,\kappa^2+
  \left(\xi\,R_1 \right)^2\,\left(3-\xi\,\tau \right) -\kappa\,\left(\xi\,R_1
  \right)^2\,(6+\tau\,\xi)\right\}.
\]
Using   (\ref{sp_0}),   we get after some algebraic manipulation the
following formula:
\[
Re \ C_1=\frac{\kappa}{16\,\Omega^2\,R_1^2}\left\{
2\,\kappa\,\xi\,\tau\,\left(\xi\,R_1
\right)^2-\chi\,\tau\,\left(\xi^2R_1
 \right)^2-3\,\left(\chi\,\xi  \right)^2\right\}.
\]
Since for $\xi=\xi_{cr}<0$ expression in braces is negative,
 the following statement is true:

\vspace{3mm}

{\bf Lemma 2.} If $R_1<R_2 $ then in vicinity of the critical value
$\xi=\xi_{cr}$ given by the formula (\ref{xicr}) a stable limit
cycle appears in system (\ref{DS_rel}).

\vspace{3mm}

We've formulated conditions assuring the appearance of periodic
orbit in proximity of stationary point $A(R_1,\,\,\Pi_1)$ yet in
order that the required homoclinic bifurcation would ever take
place, another condition should be fulfilled, namely that on the
same restrictions upon the parameters critical point
$B(R_2,\,\,\Pi_2)$ is a saddle. Besides, it is necessary to pose the
conditions on the parameters assuring that the stationary point
$B(R_2,\,\Pi_2)$ lies in the first quadrant of the phase plane for
otherwise corresponding stationary solution which is needed to to
compose the compacton would not have the physical interpretation.
Below we formulate the statement addressing both of these questions.

\vspace{3mm}

{\bf Lemma 3.} Stationary point $B(R_2,\,\,\Pi_2)$ is a saddle lying
in the first quadrant for any $\xi>\xi_{cr}$ if the following
inequalities hold:
 \begin{equation}\label{compexist}
-\tau\,\xi_{cr}\,R_2\,<\,R_1\,<\,R_2.
 \end{equation}

\vspace{3mm}

{\it Proof. } First we are going to show  that the eigenvalues
$\lambda_{1,2}$ of the system's (\ref{DS_rel})  Jacobi matrix
\begin{equation}\label{Jacobi_R2}
\hat M=\frac{\partial\left(F_1,\,F_2 \right)}{\partial\left(R,\,\Pi
\right)}\left|_{R_2,\,\Pi_2}=\left[
\begin{array}{cc}\kappa, & \frac{\sigma\,\chi}{\tau\,\xi^2} \\
\frac{\xi^2\left[\kappa(\sigma-2)+2\gamma\,R_2(\sigma-1)\right]}{\sigma},
& -\frac{\chi\sigma}{\tau}
\end{array}
\right]\right.
\end{equation}
are real and have different signs. Since the eigenvalues of $\hat M$
are expressed by the formula
\[
\lambda_{1,2}=\frac{\mbox{sp}\,\, {\hat M} \pm \sqrt{\left[
\mbox{sp}\,\, {\hat M}\right]^2-4\,\det{\hat M} }}{2},
\]
it is sufficient to show that
\begin{equation}\label{signdet}
\det{\hat M}\,<\,0.
\end{equation}
In fact, we have
\begin{eqnarray*}
\det{\hat
M}=-\frac{\chi\,\sigma\,\kappa}{\tau}-\frac{\chi}{\tau}\left[\kappa\,(\sigma-2)+2\,\gamma\,R_2\,(\sigma-1)
\right]=  \\
=-\frac{\chi}{\tau}2\,\gamma\,\tau\,\xi\left(\frac{\kappa}{\gamma}+R_2
\right)=2\,\chi\,\xi\,\gamma\,\left(R_1-R_2  \right)\,<\,0
\end{eqnarray*}

To finish the proof, we must show that stationary point
$B(R_2,\,\,\Pi_2)$ lies in the first quadrant. This is equivalent to
the statement that
\[
\kappa-\tau\,\xi_{cr}\,\gamma\,R_2>0.
\]
Carrying the first term into the RHS and dividing the inequality
obtained by $\gamma\,<\,0,$ we get the inequality
$-\tau\,\xi_{cr}\,R_2\,<\,R_1.$ The latter implies inequalities
$-\tau\,\xi\,R_2\,<\,R_1\,<\,R_2$ which are valid for any
$\xi\,>\,\xi_{cr}.$ And this ends the proof.

\begin{figure}
\centering \includegraphics[width=4.3 in, height=2.3 in
]{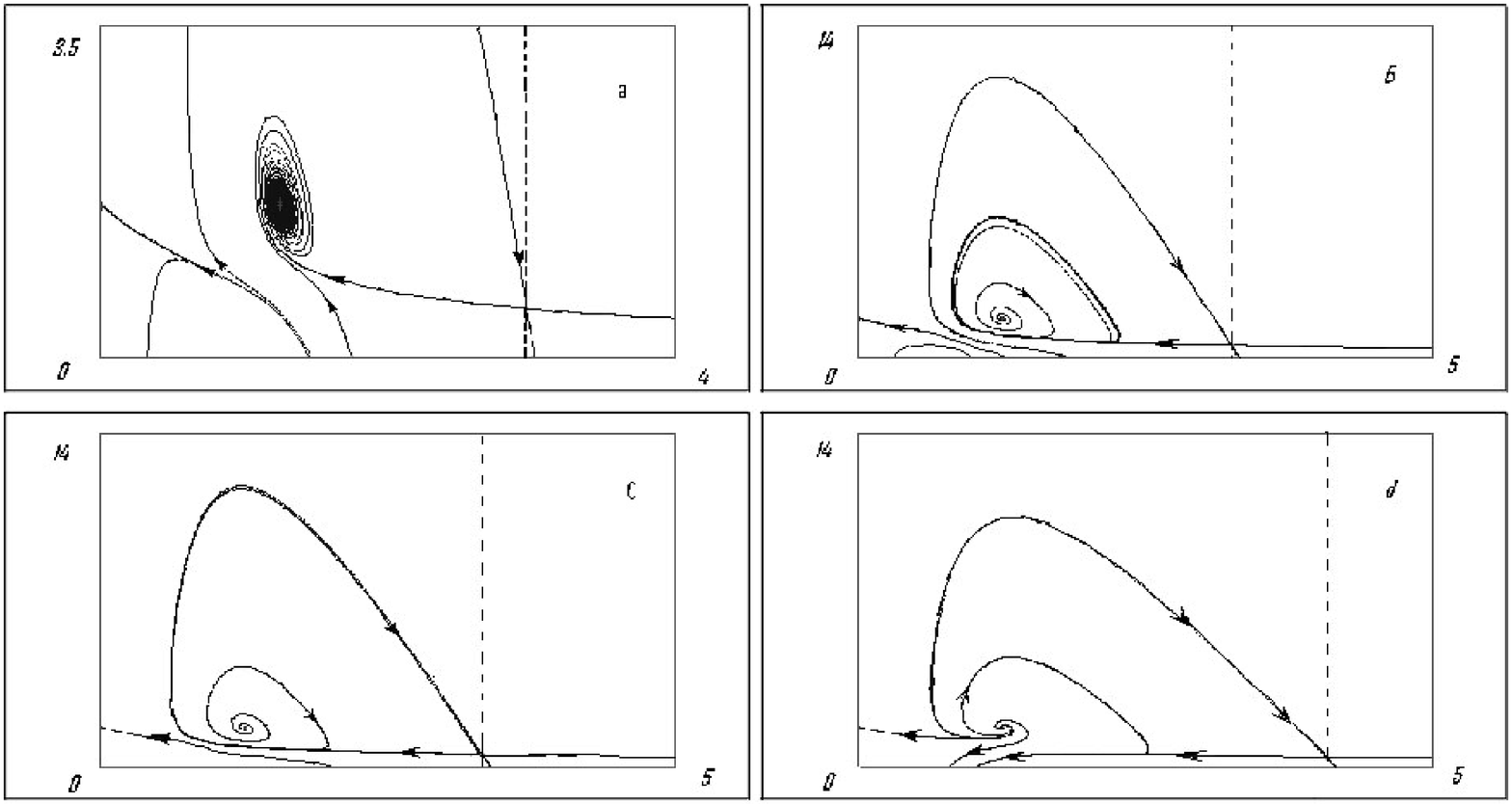} \caption{Changes of phase portrait of system
(\ref{DS_rel}): (a) $A(R_1,\,\Pi_1)$ is the stable focus; (b)
$A(R_1,\,\Pi_1)$ is surrounded by the stable limit cycle; (c)
$A(R_1,\,\Pi_1)$ is surrounded by the homoclinic loop; (d)
$A(R_1,\,\Pi_1)$ is the unstable focus;}\label{Bifpat}
\end{figure}

Numerical studies of system's (\ref{DS_rel}) behavior reveal the
following changes of regimes (cf. Fig.~\ref{Bifpat}). When $\xi
\,<\xi _{cr}$,  $A(R_1 , \Pi_1)$ is  a stable focus; above the
critical value a stable limit cycle softly  appears. Its radius
grows with the growth of parameter $\xi$ until it gains the second
critical value $\xi _{cr_2}\,>\, \xi _{cr}$ for which the homoclinic
loop appears in place of the periodic trajectory. The homoclinic
trajectory is based upon the stationary point $B(R_2 , \Pi_2)$ lying
on the line of singular points $\Delta(R)=0$ so it corresponds to
the generalized compacton-like solution to  system (\ref{Relhydro}).
We obtain this solution sewing up the TW solution corresponding to
homoclinic loop with stationary inhomogeneous solution
\begin{equation}\label{statsol}
u=0, \quad p=\Pi_2\,(x_0-x), \quad V=R_2/(x_0-x),
\end{equation}
corresponding to  critical point $B(R_2 , \Pi_2)$. So, strictly
speaking it is different from the "true" compacton, which is defined
as a solution with compact support. Note, that we can pass to the
compactly supported function by the following change of variables:
\[
\pi(t,\,x)=p(t,x)-\Pi_2\,(x_0-x), \qquad
\nu(t,\,x)=V(t,x)-R_2/(x_0-x).
\]

\section{Numerical investigations of system (\ref{Relhydro})}

\subsection{Construction and verification of the numerical scheme.}
We construct the  numerical scheme basing on the S.K.Godunov method
\cite{RozhdYan}. Since the inhomogeneous terms appearing in
(\ref{Relhydro}) destroy the scaling invariance, we look, in
accordance with common practice \cite{Romensk}, for the solution of
the Riemann problem $(V_1 u_1 p_1)$ at $x<0$ è $(V_2, u_2, p_2)$ at
$x>0$ to corresponding homogeneous system
\begin{eqnarray}\label{homogsys}
  u_t +p_x =0,\nonumber
\\
  V_t -u_x =0, \\
  p_t +\frac{\chi}{\tau V^2}V_t =0. \nonumber
\end{eqnarray}
Using the acoustic approximation, we find the functions $U,\,P$ in
the sector $-Ct<x<Ct, \ C=\sqrt{\chi /(\tau V_0^2)}$:
\begin{eqnarray}\label{acous_appr}
  U=\frac{u_1 +u_2}{2}+\frac{p_1 -p_2}{2C},
\\  P=\frac{p_1 +p_2}{2}+C \frac{u_1 -u_2}{2}. \nonumber
\end{eqnarray}
where $V_0=\frac{V_1+V_2}{2}.$ Expression for the function $V$ is
omitted since it does not take part in the construction of the
scheme on this step.

At some additional assumption the Riemann problem can be solved
without resorting to the acoustic approximation. Let us assume that
\begin{equation}\label{frozenvel}
p=\frac{\chi}{\tau V}.
\end{equation}
 As it easily seen, this relation is the
particular integral of the third equation of system
(\ref{homogsys}). Employing this formula, we can write down the
first two equations as the following closed system:

\begin{equation}\label{homo12}
\left( \frac{\partial}{\partial t}+\tilde A
\frac{\partial}{\partial x}
  \right)
  \left( \begin{array}{c}
  u \\
  V
  \end{array} \right)=0, \quad \mbox{where}\,\,\,\,
  \tilde A =
  \left( \begin{array}{cc}
  0, & -{\chi}/{(\tau V^2)} \\
  -1, & 0
  \end{array} \right).
\end{equation}
Solving the eigenvalue problem $det \ ||\tilde A -\lambda I||=0$, we
find that the characteristic velocities satisfy the equation
$$
  \lambda^2\,=\,C_{L\infty}^2\,=\,{\chi /(\tau V^2)}.
$$
Now we look for the Riemann invariants in the form of infinite
series
$$ r_{\pm}=V\sum_{\nu =0}^{\infty} A_\nu^\pm u^\nu .
$$
It is not diffiecult to verify by direct inspection that the
following relations hold:
\begin{eqnarray}\label{char_rel}
  D_{\pm}V=\left( \frac{\partial}{\partial t}
\pm C_{L\infty}\frac{\partial}{\partial x}
  \right) V=u_x \pm C_{L\infty}V_x =Q_\pm ,
\\  D_\pm u=\pm C_{L\infty}Q_\pm .
\end{eqnarray}
Using (\ref{char_rel}), we find the recurrency
$$
  A_n^\pm =(\mp 1)^n \frac{A_0}{n! (\sqrt{\chi /\tau})^n}.
$$
and finally obtain the expression for Riemann invariants:
\begin{equation}\label{Riem_inv}
  r^\pm =A_0 V \exp{ \ (\mp u / \sqrt{\chi /\tau} )}.
\end{equation}
So under the assumption that $p=\frac{\chi}{\tau V},$  system
(\ref{homogsys}) can be rewritten in the following form:
\begin{equation}\label{charform}
 D_\pm r^\pm =0.
\end{equation}
Using (\ref{frozenvel}) and (\ref{charform}) we get the solution of
Riemann problem in the sector $\sqrt{\chi /(\tau V_1^2 )}t < x <
\sqrt{\chi /(\tau V_2^2 )}t$:
\begin{eqnarray}\label{full_riemann}
  U=\sqrt{{\chi}/{\tau}} \ ln \ Z, \\
  P=p_2 +\sqrt{{\chi}/{\tau}} C_2 [Z \exp{(-u_2 /\sqrt{\chi
  /\tau})}-1], \nonumber
\end{eqnarray}
where $C_i =\sqrt{\chi /\tau}/V_i\,\equiv\, C_{L\infty}(V_i), \
i=1,2$,
$$
  Z=(E+\sqrt{Q})/(2C_2 \sqrt{\chi /\tau}),
$$
$$
  E=\exp{(u_2 /\sqrt{\chi /\tau})}\{ p_1 -p_2 +\sqrt{\chi /\tau}(C_2 -C_1)\},
$$
$$
  Q=E^2 +4 \chi /\tau C_1 C_2 \exp{[(u_1 +u_2)/\sqrt{\chi /\tau}]}.
$$
Note that (\ref{full_riemann}) is reduced to (\ref{acous_appr}) when
$|p_1 -p_2 |<<1, \ |u_1 -u_2 | <<1$.

The difference scheme for (\ref{Relhydro}) takes the following form:
\begin{eqnarray*}
(u_i^n -u_i^{n+1})\Delta x-(p_{i+1/2}^n -p_{i-1/2}^n)\Delta
t=-\gamma
  \Delta t \Delta x,
\\
(V_i^n -V_i^{n+1})\Delta x+(u_{i+1/2}^n -u_{i-1/2}^n)\Delta t=0, \\
\left( p_i^n -\frac{\chi}{\tau V_i^n}\right) \Delta x-\left(
  p_i^{n+1}-\frac{\chi}{\tau V_i^{n+1}}\right) \Delta x=-f\Delta t\Delta x,
\end{eqnarray*}
where $(u_{i-1/2}^n , \, p_{i-1/2}^n)$ è $(u_{i+1/2}^n , \,
p_{i+1/2}^n)$  are solutions of Riemann problems $(V_{i-1}^n, \,
u_{i-1}^n, \, p_{i-1}^n),$ $(V_i^n, \, u_i^n, \, p_i^n)$ è $(V_i^n,
\, u_i^n, \, p_i^n)$, $(V_{i+1}^n, \, u_{i+1}^n, \, p_{i+1}^n)$,
correspondingly,
$$
  f=f(p_i^k, \, V_i^k)=\frac{\kappa}{V_i^k}-p_i^k,
$$
$k$ is equal to either $n$ or $n+1$. The choice $k=n$ leads to the
explicit Godunov scheme
\begin{equation}\label{expl_Godunov}
  \left\{ \begin{array}{c}
  u_i^{n+1}=u_i^n +\frac{\Delta t}{\Delta x}\left(p_{i-1/2}^n-p_{i+1/2}^n\right)
  + \gamma \Delta t \\ V_i^{n+1}=V_i^n +\frac{\Delta
  t}{\Delta x}(u_{i+1/2}^n -u_{i-1/2}^n) \\ p_i^{n+1}=p_i^n
  +\frac{\chi}{\tau}\left( \frac{1}{V_i^{n+1}}-\frac{1} {V_i^n}\right)
  +f(p_i^n, \, V_i^n)\Delta t \end{array} \right.
  \end{equation}

The scheme (\ref{expl_Godunov}) was tested on invariant TV solutions
of the following form:
\begin{equation}\label{test_self}
  u=U(\omega), \quad p=P(\omega), \quad V=V(\omega), \quad \omega =x-Dt.
\end{equation}
Inserting (\ref{test_self}) into first two equations of system
(\ref{Relhydro}), one can obtain the following first integrals:
\begin{equation}\label{firstint}
  \left\{ \begin{array}{c}
  U=u_1 +D(V_1 -V), \\
  P=p_1 +D^2 (V_1 -V),
  \end{array} \right.
\end{equation}
where $V_1 =\ lim_{\omega\to\infty}\, V(\omega)$. Let us assume in
addition that $u_1 =0$, while  $\ p_1 =\kappa /V_1$. With this
assumption constants $(u_1, \, p_1, \, V_1)$ satisfy the initial
system.

Inserting $U\,\,\mbox{and}\,\,\,P$ into the third equation of system
(\ref{Relhydro}) we get:
\begin{equation}\label{eqfor_V}
\frac{dV}{d\omega}=-V\frac{[D^2 V^2 -SV+\kappa]} {\tau
D[C_{T\infty}^2 -(DV)^2]}\,=\,F(V),
\end{equation}
where $C_{T\infty} =\sqrt{\chi /\tau}, \ S=p_1 +D^2 V_1$. Equation
(\ref{eqfor_V}) has three critical points:
$$
  V=V_0=0, \qquad V=V_1, \qquad V=V_2=\kappa /(V_1 D^2).
$$

\begin{figure}
\centering \includegraphics[width=3.0 in, height=2.6 in ]{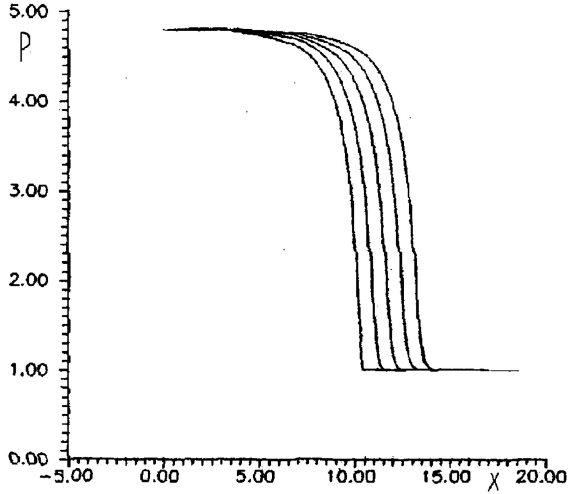}
\caption{Temporal evolution of Cauch\'{y} data defined by solutions
of equation (\ref{eqfor_V}) and the first integrals
(\ref{firstint}). Following vaues of the parameters were chosen
during the numerical simulation:
$\kappa\,=\,0.5,\,\,\chi\,=\,0.25,\,\,\tau\,=\,0.1,\,\,
V_1\,=\,0.5,\,\,D\,=\,3.1.$ } \label{Test}
\end{figure}

If the inequality $V_2 =\kappa /(V_1 D^2) <V_1$ holds and the line
$\chi/\tau -(DV)^2 =0$ is outside the interval then $(V_2, \, V_1)$
then constants
$$
  u_{-\infty}=u_2 =\frac{(DV_1)^2 -\kappa}{DV_1}>0, \qquad
  p_{-\infty}=p_2 =\kappa /V_2 , \qquad V_{-\infty}=V_2,
$$
deliver the second stationary solution to the initial system and
solution to (\ref{eqfor_V}) corresponds to a smooth compressive wave
connecting these two stationary solutions.

Results of numerical solving the Cauch\'{y} problem  based on the
Godunov scheme (\ref{expl_Godunov}) are shown in Fig.~\ref{Test}. As
the Cauch\'{y} data we took the smooth self-similar solution
obtained by numerical solving equation (\ref{eqfor_V}) and
employment of the first integrals (\ref{firstint}). So we see that
the numerical scheme quite well describes the self-similar evolution
of the initial data.

\subsection{Numerical investigations of the temporal evolution and attractive features of compactons.}

Below we present the results of numerical solving of the Cauch\'{y}
problem for system (\ref{Relhydro}). In numerical experiments  we
used the values of the parameters taken in accordance with the
preliminary results of qualitative investigations and corresponding
to the homoclinic loop appearance in system (\ref{DS_rel}).  As the
Cauch\'{y} data we got the generalized solution describing the
compacton and obtained by the preliminary solving of system
(\ref{DS_rel}) and employment of formulae (\ref{quadr}),
(\ref{statsol}). Results of the numerical simulation are shown in
Fig.~\ref{Compevol}. It is seen that compacton evolves for a long
time in a stable self-similar mode.

Additionally the numerical experiments revealed that the wave packs
being created by sufficiently wide family of initial data tend under
certain conditions to the compacton solution. Following family of
the initial perturbations have been considered in the numerical
experiments:

\begin{figure}
 \centering\includegraphics[width=2.5 in, height=2 in ]{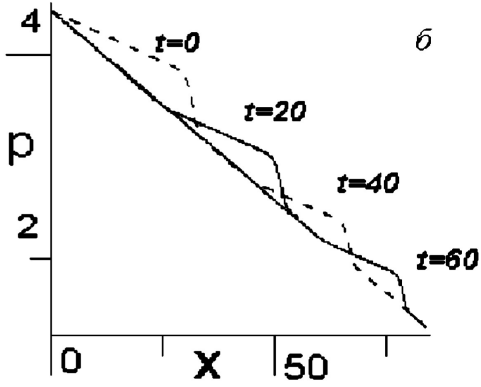}
\caption{Numerical solution of system (\ref{Relhydro}) in case when
the invariant  homoclinic solution is taken as the Cauch\'{y}
data}\label{Compevol}
\end{figure}
\begin{equation}\label{initpert}
p\,\,=\left\{
\begin{array}{c}
p_{0}(x_0-x) \quad \mbox{when} \quad x \in (0,a) \cup (a+l,x_0) \\
(p_{0} +p_{1})(x_0-x)+ w(x-a)+h  \quad\mbox{when} \quad x \in (a,
a+l),
\\ u=0, \qquad V=\kappa/p.
\end{array} \right.
 \end{equation}
Here $a,\,l,\,p_1,\,w,\,h$  are parameters of  the perturbation
defined on the background of the inhomogeneous stationary solution
(\ref{statsol}). Note that parameter $l$ defines the width of the
initial perturbation. Varying broadly parameters of the initial
perturbation, we observed in numerical experiments that,  when
fixing e.g. value of $l$, it is possible to fit in many ways the
rest of parameters such that one of the wave packs created by the
perturbation (namely that one which runs "downwards"  towards the
direction of diminishing pressure) in the long run approaches
compacton solution. Whether the wave pack would approach the
compacton solution or not occurs to depend on that part of energy of
the initial perturbation which is carried out "downwards". Assuming
that the energy is divided between two wave packs created more or
less in half, we can use for the rough estimation of convergency the
total energy of the initial perturbation, consisting of the internal
energy $E_{int}$ and the potential energy $E_{pot}$:

\begin{figure}
\includegraphics[width=2.0 in, height=1.9 in ]{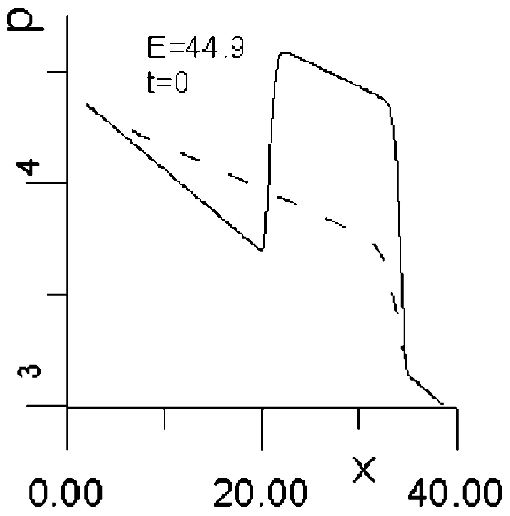}
\hfill
\includegraphics[width=2.0 in, height=1.9 in ]{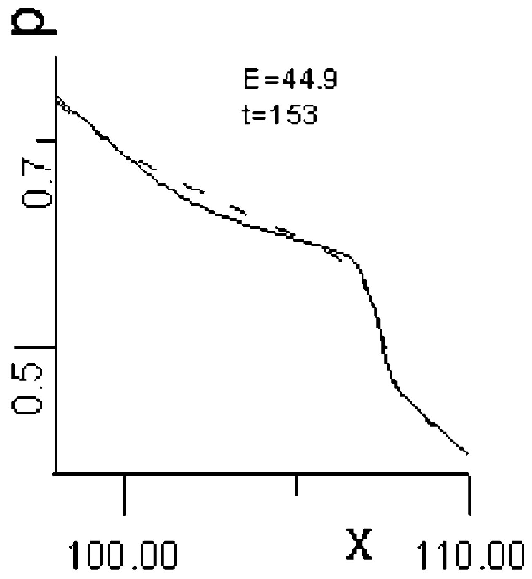}
\includegraphics[width=2.0 in, height=1.9 in  ]{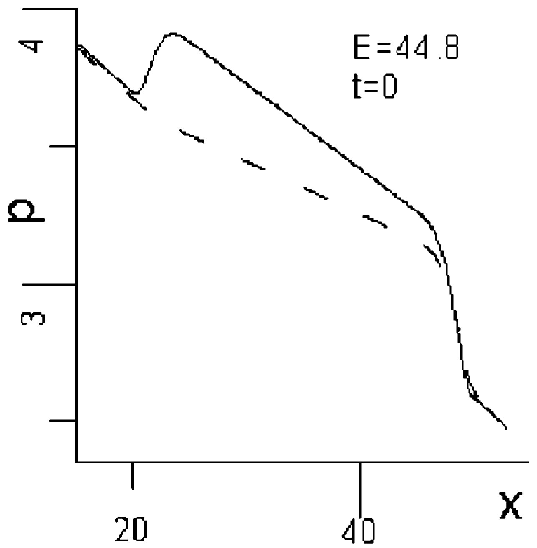}
\hfill
\includegraphics[width=2.0 in, height=1.9 in ]{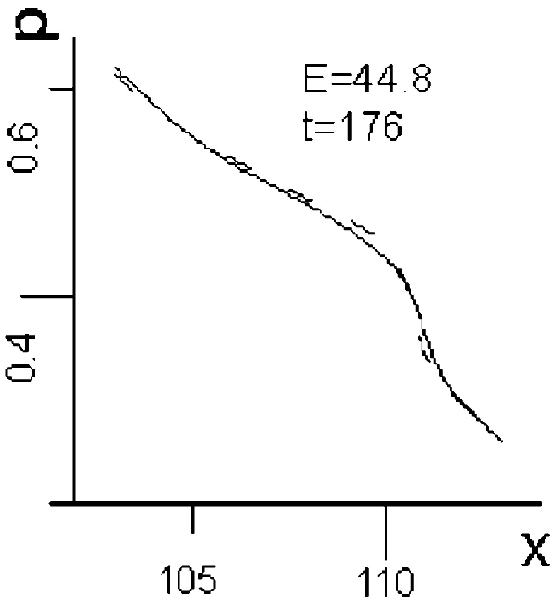}
\includegraphics[width=2.0 in, height=1.9 in ]{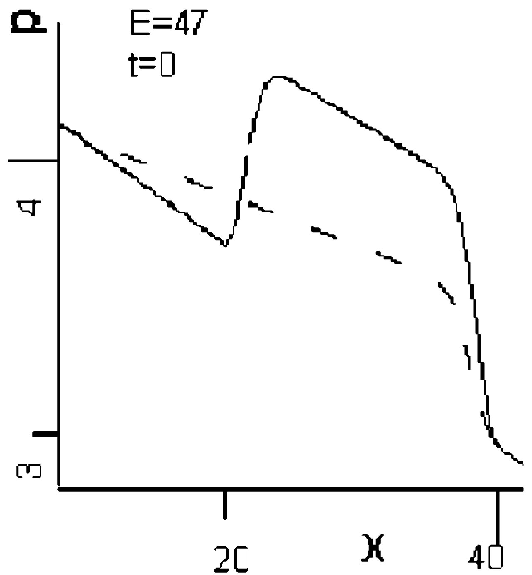}
\hfill
\includegraphics[width=2.0 in, height=1.9 in ]{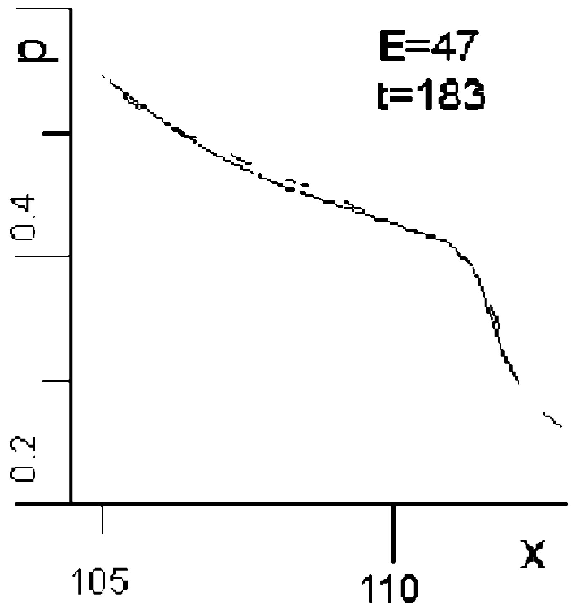}
\caption{Perturbations of ÿstationary invariant solutions of system
(\ref{Relhydro}) (left) and TW soluations created by these
perturbations (right) on the background of the invariant
compacton-like solution (dashed) }\label{Attractors}
\end{figure}

$$
E=E_{int}+E_{pot}= \int \left[ \varepsilon_{int}\,
+\varepsilon_{pot}\right]dx,
$$
where $\varepsilon_{int},\,\,\mbox{and}\,\,\,\varepsilon_{pot}\,$
are local densities of the corresponding terms

Function $\varepsilon_{pot}$ is connected with forces acting in
system by means of  the evident relation
$$
\gamma \,-  \frac{1}{\rho}\frac{\partial p}{\partial x_e} \,=
\,-\frac{\partial\varepsilon_{pot}}{\partial x_e},
$$
where $x_e$ is physical (Eulerian) coordinate connected with the
mass Lagrangean coordinate $x$ as follows:
\[
x=\int{V^{-1}\,d\,x_e}.
\]
From this we extract the expression
$$
E_{pot}=\int \varepsilon_{pot\,}d\,x\,=\int\limits_{\Omega }\left[
\int\limits_{c_1}^{x_e}\left(V\frac{\partial p}{\partial x_e}
-\gamma \right)dx_e^{\prime}\right] V^{-1} dx_e,
$$
where $\Omega$  is the support of initial perturbation.Employing in
the above integral the relation $V\,\partial p/\partial
x_e\,=\partial p/\partial x,$ we obtain the following formula:
$$ E_{pot}\,= \frac{\kappa\, l}{\alpha +\beta}
\left[\biggl(\frac{1+k}{k}\biggr)\ln (1\,+k)\,-1\right] ,
$$
where $k =[ P\,(a\,+l)- P\,(a)]/ P\,(a), \quad P(z)\,=\alpha
z\,+\,\beta ,\quad  \alpha \,=w-(p_{0} +p_1 ),\,\quad
\beta\,~=(p_{1}+p_0)x_0\,-aw.$

For $\chi\,=\,1.5$, $\kappa\,=\,10$, $\gamma\,=\,-0.04$,
$\tau\,=\,0.07$, $x_0\,=\,120$ convergency was observed when
$E_{tot}$ was close to 45 (see Figures below).

\begin{figure}
\includegraphics[width=2.1 in, height=2.0 in ]{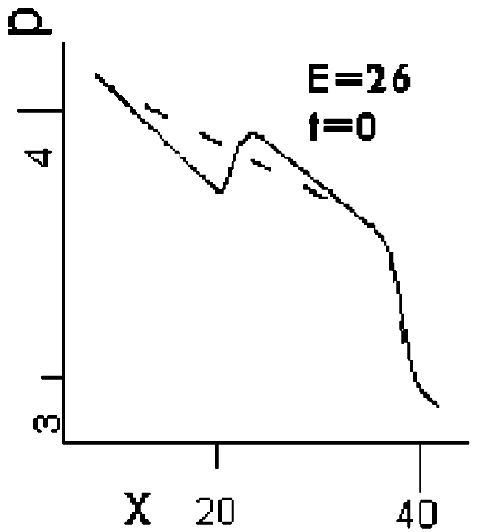}
\hfill
\includegraphics[width=2.1 in, height=2.0 in]{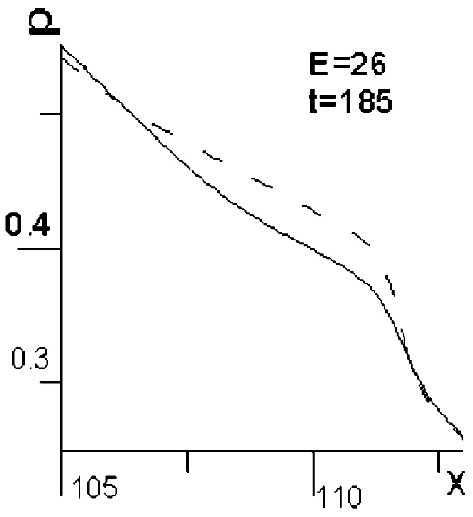}
\includegraphics[width=2.1 in, height=2.0 in ]{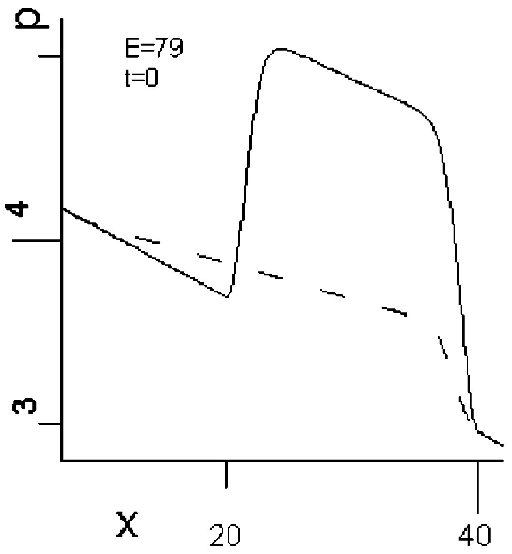}
\hfill
\includegraphics[width=2.1 in, height=2.0 in]{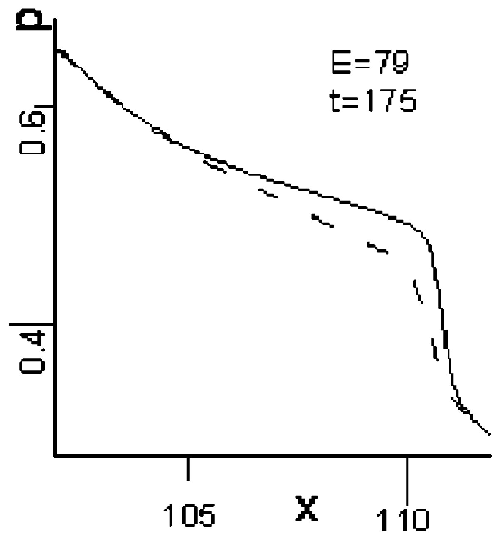}
\caption{Evolution of the wave patterns created by the local
perturbations which do not satisfy the energy criterium
}\label{Repellers}
\end{figure}

Function $\varepsilon_{int}$ is obtained from the second low of
thermodynamics written for the adiabatic case: $(\partial
\varepsilon_{int}/\partial V)_S\,=-p\,=-\kappa /V$. From this we get
$$
\varepsilon_{int}\,=c\,-\kappa \ln V.
$$
To obtain the energy of perturbation itself, we should subtract from
this value the energy density of stationary inhomogeneous solution
$c-\kappa \ln\,V_{0}$, so finally we get
$$
E_{int}\,=\int\limits_{\Omega }(\varepsilon_{int}\,-
\varepsilon_{int}^0)V^{-1} dx_e\,\,= \kappa \int\limits_{\Omega }\ln
{V_0/V} dx_{l}.
$$
Using the forula (3.7.20),we finally obtain
$$
E_{int}=\kappa\left\{ l\ln\frac{ P(a\,+l)}{P_0(a\,+l)} +\frac{
P(a)}{\alpha} \ln \left[ 1+\frac{\alpha l}{P(a)}\right]
+\frac{P_0(a)}{p_2}\ln\left[ 1\,-\frac{p_2 l}{P_0(a)}\right]
\right\},
$$
where $P_0 (z)\,=p_0 (x_0 -z).$

Numerical experimenting shows that the energy norm serves as
sufficiently good criterium of convergency. At $\chi\,=\,1.5$,
$\kappa\,=\,10$, $\gamma\,=\,-0.04$, $\tau\,=\,0.07$, $x_0\,=\,120$
convergency was observed when $E\in (43,47).$ Patterns of evolution
of the wave perturbations is shown in Figs~\ref{Attractors}. For
comparison we also show the temporal evolution of the wave packs
created by the perturbations for which $E\not\in (43,\,\,47)$
(Figs.~\ref{Repellers}).

Thus there is observed some correlation between the energy of
initial perturbation and convergency of the wave packs created to
the compacton solution.

\section{Conclusions and discussion.}

In this work we have discussed the origin of generalized TW
solutions called compactons and  have shown the existence of such
solutions within the hyd\-ro\-dy\-na\-mic-type model of relaxing
media. The main results concerning this subject can be summarized as
follows:
\begin{itemize}
\item
A family of TW solutions to (\ref{Relhydro}) includes  a compacton
{\bf  in case when an external force is present (more precisely,
when $\gamma<0$.)}
\item
Compacton solution to system (\ref{Relhydro}) occurs merely at
selected values of the parameters: {\bf for fixed
$\kappa,\,\,\gamma\,\,\mbox{and}\,\,\chi$ there is a unique
compacton-like solution, corresponding to the value
$\xi=\xi_{cr_2}$}.
\end{itemize}

 Qualitative an numerical analysis of the corresponding
ODE system describing the TW solutions to initial system served us
as a starting point in numerical investigations of compactons, based
on the Godunov method. Numerical investigations reveal that
compacton encountering in this particular model form a stable wave
pattern evolving in a self-similar mode. It is also obtained a
numerical evidence of attracting features of this structure: a wide
class of initial perturbations creates wave packs tending to
compacton. Convergency only weakly depend on the shape of initial
perturbation and is mainly caused by fulfillment of the energy
criterion. This criterium is far from being perfect. In fact, it is
not sensible on the form of initial perturbation, which, in turn,
influences the part of the  the total energy getting away by the
wave pack moving "downwards". Besides, employment of the Godunov
scheme does not enable to obtain more strict quantitative measure of
convergency. But in spite of these discrepancies the effect of
convergency is evidently observed and this will be the topic of our
further study to develop more strict criteria of convergency as well
as trying to realize whether the compacton solution serves as true
or intermediate \cite{Barenblatt,Bedlewo} asymptotics.

\end{document}